\documentclass{article}
\usepackage{spconf,amsmath,graphicx,amssymb,xcolor,color,dsfont}
\usepackage[hidelinks]{hyperref}
\usepackage{booktabs, float, verbatim}
\usepackage{cite}

\newcommand{\Mu}{\boldsymbol{\mu}}
\newcommand{\Eta}{\boldsymbol{\eta}}
\newcommand{\x}{\boldsymbol{x}}
\newcommand{\transpose}{^{^{\intercal}}}
\newcommand{\inv}{^{^{-1}}}

\usepackage{booktabs}
\usepackage{multirow}
\usepackage{tabularx}
\newcolumntype{Y}{>{\centering\arraybackslash}X}
\usepackage{lipsum}

\newcommand\blfootnote[1]{%
  \begingroup
  \renewcommand\thefootnote{}\footnote{#1}%
  \addtocounter{footnote}{-1}%
  \endgroup
}

\title{Pairwise Discriminative Neural PLDA for Speaker Verification}
\name{{Shreyas Ramoji, Prashant Krishnan V, Prachi Singh, Sriram Ganapathy}}  
\address{Learning and Extraction of Acoustic Patterns (LEAP) Lab \\ Department of Electrical Engineering, Indian Institute of Science, Bengaluru}

\ninept

\begin{document}

\maketitle
\begin{abstract}
The state-of-art approach to speaker verification involves the extraction of discriminative embeddings like x-vectors followed by a generative model back-end using a probabilistic linear discriminant analysis (PLDA). In this paper, we propose a Pairwise neural discriminative model for the task of speaker verification which operates on a pair of speaker embeddings such as x-vectors/i-vectors and outputs a score that can be considered as a scaled log-likelihood ratio. We construct a differentiable cost function which approximates speaker verification loss, namely the minimum detection cost. The pre-processing steps of linear discriminant analysis (LDA), unit length normalization and within class covariance normalization are all modeled as layers of a neural model and the speaker verification cost functions can be back-propagated through these layers during training. We also explore regularization techniques to prevent overfitting, which is a major concern in using discriminative back-end models for verification tasks. The experiments are performed on the  NIST SRE 2018 development and evaluation datasets. We observe average relative improvements of 8\% in CMN2 condition and 30\% in VAST condition over the PLDA baseline system. 
\end{abstract}


\section{Introduction}\label{sec:Intro}
The earliest successful approach to speaker recognition used the Gaussian mixture modeling (GMM) from the training data followed by an adaptation using maximum-aposteriori (MAP) rule \cite{reynolds2000speaker}. The development of i-vectors as fixed dimensional front-end features for speaker recognition tasks was introduced in \cite{kenny2007joint,dehak2011front}.  Recently, neural network embeddings trained on a  speaker discrimination task were also derived as features to replace the i-vectors. These features called x-vectors \cite{snyder2018x} were shown to perform better than the i-vectors for speaker recognition~\cite{mclaren2018train}. 

Following the extraction of x-vectors/i-vectors, different pre-processing steps are employed to transform the embeddings. The common steps include linear discriminant analysis (LDA)~\cite{dehak2011front}, unit length normalization~\cite{garcia2011analysis} and within-class covariance normalization (WCCN)~\cite{hatch2006within}. The transformed vectors are modeled with probabilistic linear discriminant analysis (PLDA)~\cite{kenny2010bayesian}. The PLDA model is used to compute a log likelihood ratio from a pair of enrollment and test embeddings which is used to verify whether the given trial is a target or non-target. 

In this paper, we propose a neural back-end model which jointly performs pre-processing and scoring. It operates on pairs of x-vector embeddings (a pair of enrollment and test x-vectors), and outputs a score that allows the decision of target versus non-target hypotheses. 
The implementation using neural layers allows the entire model to be learnt using a speaker verification cost. The use of conventional cost functions like binary cross entropy tend to overfit the model to the training speakers, thereby performing poorly on evaluation sets. In an attempt to avoid this, we use the NIST SRE normalized detection cost \cite{sre18evalplan} to optimize the neural back-end model. With several experiments on the NIST SRE 2018 development and evaluation dataset, we show that the proposed approach improves significantly over the state-of-the-art x-vector based PLDA system. 



The rest of the paper is organized as follows. In Section~\ref{sec:related_work}, we highlight relevant prior work done in the field of discriminative back-end for speaker verification.  Section \ref{sec:fe_feature_xvector} describes the front-end configurations used for feature processing and x-vector extraction. Section \ref{sec:PldaNet} describes the proposed neural network architecture used, and the connection with generative PLDA model. In Section \ref{sec:costfuncs}, we present a smooth approximation to the NIST SRE detection cost function, and discuss regularization methods.  This is followed by discussion of results in Section \ref{sec:results} and a brief set of concluding remarks in Section \ref{sec:summary}.

\section{Related Prior Work}
\label{sec:related_work}
 The common approaches for scoring in speaker verification systems  include support vector machines (SVMs) \cite{campbell2006support}, Gaussian back-end model \cite{mclaren2013adaptive,benzeghiba2009language} and the probabilistic linear discriminant analysis (PLDA) \cite{kenny2010bayesian}. Some efforts on pairwise generative and discriminative modeling are discussed in \cite{cumani2013pairwise,cumani2014large,cumani2014generative}. The discriminative version of PLDA with logistic regression and support vector machine (SVM) kernels has also been explored in ~\cite{burget2011discriminatively}. In this work, the authors use the functional form of the generative model and pool all the parameters needed to be trained into a single long vector. These parameters are then discriminatively trained using the SVM loss function with pairs of input vectors. The discriminative PLDA (DPLDA) is however prone to over-fitting on the training speakers and leads to degradation on unseen speakers in SRE evaluations~\cite{villalba2019state}. The regularization of embedding extractor network using a Gaussian back-end scoring  has been investigated in \cite{ferrer2019optimizing}. 
 
 Recently, end-to-end approaches to speaker verification have also been examined. For example, in~\cite{rohdin2018end}, the i-vector extraction with PLDA scoring has been jointly derived using a deep neural network architecture and the entire model is trained using a binary cross entropy training criterion. The use of triplet loss in end-to-end speaker recognition has shown promise for short utterances~\cite{zhang2017end}.  Wan et. al.~\cite{wan2018generalized} proposed a generalized end-to-end loss inspired by minimizing the centroid mean of within speaker distances while maximizing across speaker distances.  However, in spite of these efforts, most of the successful systems for SRE evaluations continue to use the generative PLDA back-end model. 
 
In this paper, we argue that the major issue of over-fitting in discriminative back-end systems arises from the choice of the model and loss function. In the detection cost metrics ($C_{min}$ and $C_{primary}$ ) for SRE, the false-alarm errors have more significance compared to miss errors.  Thus, incorporating the SRE evaluation metric directly in the optimization avoids the over-fitting problem. Further, by training multiple pre-processing steps along with the scoring module, the model learns to generate representations that are better optimized for the speaker verification task.

\section{Speaker Embedding Extractor}\label{sec:fe_feature_xvector}
In this section, we provide the description of the front-end feature extraction and x-vector model configuration.
\subsection{Training }
The x-vector extractor is trained entirely using speech data extracted from combined VoxCeleb 1 ~\cite{nagrani2017voxceleb} and VoxCeleb 2 corpora~\cite{chungvoxceleb2}. These datasets contain speech extracted from celebrity interview videos available on YouTube, spanning a wide range of different ethnicities, accents, professions, and ages. For training the x-vector extractor, we use $1,276,888$ segments from $7323$ speakers selected from Vox-Celeb 1 (dev and test), and VoxCeleb 2 (dev).

This x-vector extractor was trained using $23$ dimensional Mel-Frequency Cepstral Coefficients (MFCCs) from $25$ ms frames shifted every $10$ ms using a $23$-channel mel-scale filterbank spanning the frequency range $20$ Hz - $3700$ Hz. A 5-fold augmentation strategy is used that adds four corrupted copies of the original recordings to the training list~\cite{snyder2018x,mclaren2018train}. 
The augmentation step  generates
$6,384,440$ training segments for the combined VoxCeleb set.

\subsection{The x-vector extractor}
For x-vector extraction, an extended TDNN with $12$ hidden layers and rectified linear unit (RELU) non-linearities is trained to discriminate among the nearly $7000$ speakers in the training set~\cite{mclaren2018train}. The first $10$ hidden layers operate at frame-level, while the last $2$ layers operate at segment-level.   There is a $1500$-dimensional statistics pooling layer between the frame-level and segment-level layers that accumulates all frame-level  outputs using mean and standard deviation.  After training, embeddings are extracted from the $512$ dimensional affine component of the $11$th layer (i.e., the first segment-level layer).  More details regarding the DNN architecture and the training process can be found in \cite{mclaren2018train}.

\section{Pairwise Discriminative Neural PLDA Back-end}\label{sec:PldaNet}
Following the x-vector extraction, the embeddings are centered (mean removed), transformed using LDA and unit length normalized. The PLDA model on the processed x-vector for a given recording  is,
\begin{equation}
\Eta_r = \Phi \omega + \boldsymbol{\epsilon}_r 
\end{equation}
where $\Eta_r$ is the x-vector for the given recording, $\omega$ is the latent speaker factor with a prior of $\mathcal{N}(0,I)$, $\Phi$ characterizes the speaker sub-space matrix and $\boldsymbol{\epsilon}_r$ is the residual assumed to have distribution $\mathcal{N}(0,\Sigma)$. \blfootnote{The implementation of the Neural PLDA Network can be found here: \url{https://github.com/iiscleap/NeuralPlda}}.

\begin{figure}[t]
    \hspace{-1ex}\includegraphics[width=1\linewidth, trim={2.2cm 2.1cm 2cm 1.5cm},clip]{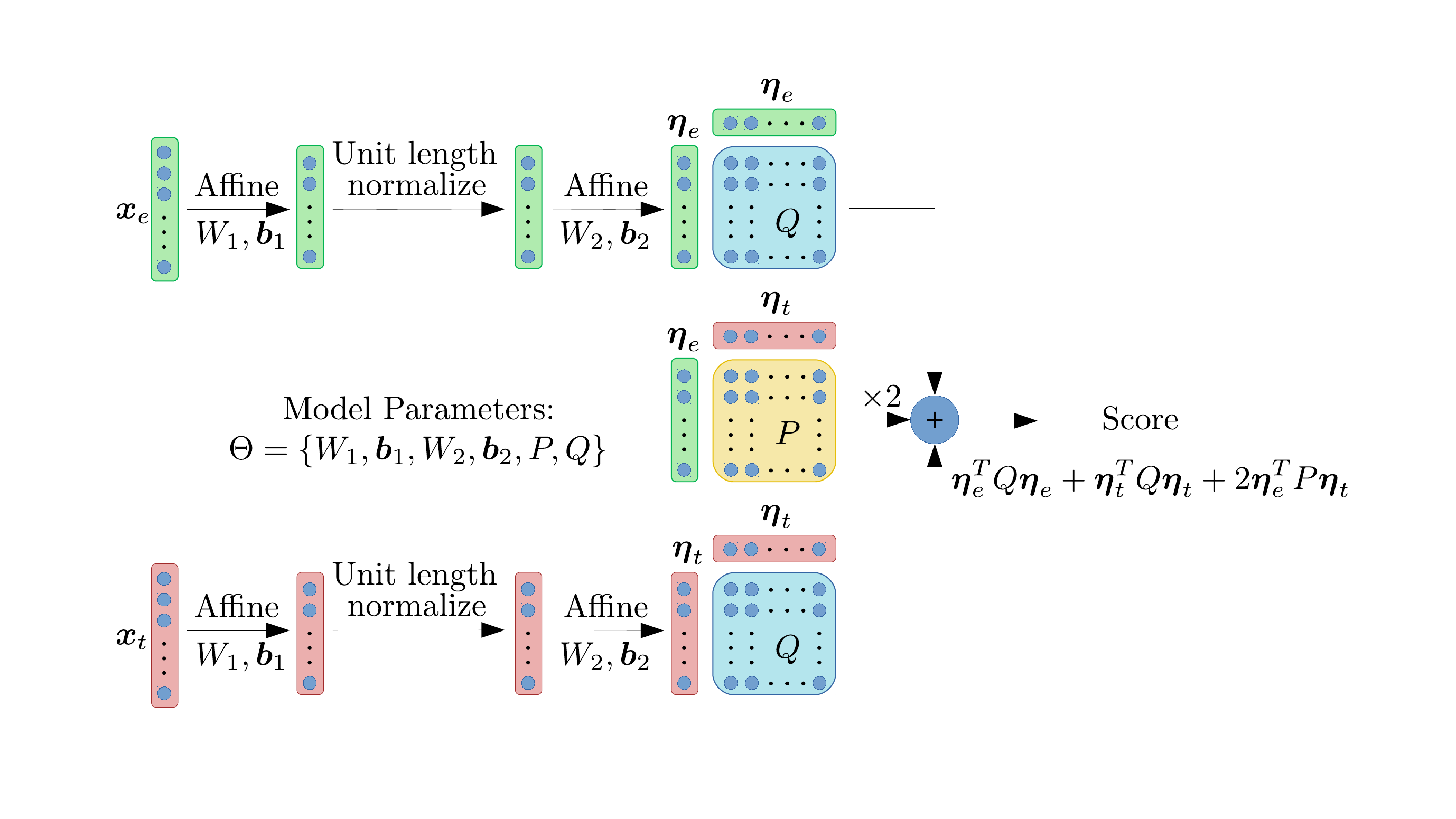}
    \caption{Neural PLDA Net Architecture: The two inputs $\x_1$ and $\x_2$ are the enrollment and test x-vectors which constitute a \textit{trial}.}
    \label{fig:PldaNet}
\end{figure}

For scoring, a pair of x-vectors, one from the enrollment recording $\Eta_e$ and one from the test recording $\Eta_t$ are used with the pre-trained PLDA model to compute the log-likelihood ratio score as,
\begin{equation}\label{eq:plda_scoring}
s(\Eta_e, \Eta_t) = \Eta_e\transpose Q \Eta_e + \Eta_t\transpose Q \Eta_t + \Eta_e\transpose P \Eta_t  
\end{equation}
where, 
\begin{eqnarray}
Q = \Sigma _{tot} ^{-1} -  (\Sigma _{tot} - \Sigma _{ac} \Sigma _{tot}^{-1} \Sigma _{ac})^{-1} \\
P =  \Sigma _{tot} ^{-1} \Sigma _{ac} (\Sigma _{tot} - \Sigma _{ac} \Sigma _{tot}^{-1} \Sigma _{ac})^{-1}
\end{eqnarray}
with $\Sigma _{tot} = \Phi \Phi \transpose + \Sigma$ and $\Sigma _{ac} = \Phi \Phi \transpose $. 
In the proposed pairwise discriminative network (Neural PLDA)  (Fig.~\ref{fig:PldaNet}), we construct the pre-processing steps of LDA as first affine layer, unit-length normalization as a non-linear activation and PLDA centering and diagonalization as another affine transformation. The final PLDA pair-wise scoring given in Eq.~\ref{eq:plda_scoring} is implemented as a quadratic layer in Fig.~\ref{fig:PldaNet}. Thus, the Neural PLDA implements the pre-processing of the x-vectors and the PLDA scoring as a neural back-end. The model parameters of the Neural PLDA can be initialized with the baseline system and these parameters can be learned in a backpropagation setting.

\section{Cost Function and Regularization}\label{sec:costfuncs}
To train the Neural PLDA for the task of speaker verification, it is required to sample pairs of x-vectors representing target (from same speaker) and non-target hypothesis (from different speakers).  We train the model using the trials from previous NIST SRE evaluation sets along with randomly sampled target and non-target pairs which are matched by source and gender.
 The following error functions can be used in the Neural PLDA,
\subsection{Binary Cross Entropy}
The standard objective for a two class classification task.
\begin{align}
    L_{BCE} = \frac{1}{N}\sum_{i=1}^{N} t_i \log \sigma(s_i) + (1-t_i) \log (1-\sigma (s_i))
\end{align}
where $s_i$ is the score for the $i^{\text{th}}$ trial, $t_i$ is the binary target for the trial and $N$ is the number of trials. 

Using this loss alone for training may result in over-fitting. Hence, a regularization term can be  used by regressing to raw PLDA scores generated from Kaldi. The regularized cross-entropy loss is given as:
\begin{align}
    L_{BCE}^{\prime} = L_{BCE} + \frac{\lambda}{N}\sum_{i=1}^{N} (s_i -l_i)^2
\end{align}
The second term encourages the scores from the Neural PLDA to not digress from the generative model PLDA scores drastically.

\subsection{Soft Detection Cost}
The NIST SRE 2018 normalized detection cost metric \cite{sre18evalplan} is defined as:
\begin{align}\label{eq:det_cost}
    C_{Norm}(\beta,\theta) = P_{Miss}(\theta) + \beta P_{FA}(\theta)
\end{align}
where $P_{Miss}$ and $P_{FA}$ are the probability of miss and false alarms computed by applying detection threshold of $\theta$,
\begin{align}
    P_{Miss}(\theta) &= \frac{\sum_{i=1}^{N} t_i \mathds{1}(s_i<\theta)}{\sum_{i=1}^{N} t_i}\\
    P_{FA}(\theta) &= \frac{\sum_{i=1}^{N} (1-t_i) \mathds{1}(s_i \geq \theta)}{\sum_{i=1}^{N} (1-t_i)}.
\end{align}
Here, $\mathds{1}$ is the indicator function. The normalized detection cost function (Eq.~\ref{eq:det_cost}) is not a smooth function of the parameters due to the step discontinuity induced by the indicator function $\mathds{1}$, and hence, it cannot be used as an objective function in a neural network. We propose a differentiable approximation of the normalized detection cost by approximating the indicator function with a sigmoid function. 
\begin{align}
    P_{Miss}^{\text{(soft)}}(\theta) &= \frac{\sum_{i=1}^{N} t_i  \left[1-\sigma(\alpha(s_i-\theta))\right]}{\sum_{i=1}^{N} t_i} \\
    P_{FA}^{\text{(soft)}}(\theta) &= \frac{\sum_{i=1}^{N} (1-t_i) \sigma(\alpha(s_i - \theta))}{\sum_{i=1}^{N} (1-t_i)}
\end{align}
By choosing a large enough value for $\alpha$, the approximation can be made arbitrarily close to the actual detection cost function for a wide range of thresholds.

The primary cost metric of the NIST SRE 2018 for the Conversational Telephone Speech (CTS) is given by
\begin{align}
    C_{Primary} = \frac{1}{2}\left[C_{Norm}(\beta_1, \log \beta_1) + C_{Norm}(\beta_2, \log \beta_2)\right]
\end{align}
where $\beta_1 = 99$ and $\beta_2 = 199$. We compute the Neural PLDA loss function as
\begin{align}
     \mathcal{L}_{Primary} = \frac{1}{2}\left[C_{Norm}^{\text{(soft)}}(\beta_1, \theta_1) + C_{Norm}^{\text{(soft)}}(\beta_2, \theta_2)\right]
\end{align}


where $\theta_1$ and $\theta_2$ are the thresholds which minimizes $\mathcal{L}_{Primary}$.  The minimum detection cost is achieved at a threshold where $C_{primary}$ is minimized. In other words, it is the best cost that can be achieved through calibration of the scores. We include these thresholds in the set of parameters that the neural network learns to minimize $C_{Min}$ through backpropagation. Finally, we compute an affine calibration transform using the SRE 2018 development set.

\begin{table*}[t!]
\centering
\caption{ Summary of results of various back-end models on CMN2 and VAST datasets reported on the SRE 2018 development and evaluation
datasets.}
\vspace{1ex}
\begin{tabularx}{0.9 \textwidth}{@{}Y|Y|Y|Y|Y|Y||Y|Y|Y@{}}
\toprule
\multicolumn{2}{c|}{\multirow{2}{*}{Systems}} & {\multirow{2}{*}{Dataset}} & \multicolumn{3}{c||}{Dev} & \multicolumn{3}{c}{Eval} \\ \cmidrule(l){4-9} 
\multicolumn{2}{c|}{} &  & EER (\%) & $C_{min}$ & $C_{primary}$ & EER (\%) & $C_{min}$ & $C_{primary}$  \\ \midrule
\multicolumn{2}{c|}{PLDA Baseline} & CMN2 &10.02 &0.583 &0.600  &11.50	&0.642	&0.675   \\
\multicolumn{2}{c|}{(Kaldi)} & VAST &11.11	& 0.605	& 0.782  &12.70	&0.686	&0.766   \\ \midrule
\multicolumn{2}{c|}{DPLDA Baseline \cite{burget2011discriminatively} } & CMN2 &11.91	&0.683	&0.718   &  13.19 & 0.732 & 0.78  \\
\multicolumn{2}{c|}{} & VAST & 11.11 & 0.527 & 0.560 &14.68	& 0.625	& 0.629   \\ \midrule
\multicolumn{2}{c|}{\multirow{2}{*}{Pairwise GB \cite{ramoji2019LEAP,cumani2014generative}}} & CMN2 &12.57	&0.606	&0.62  &12.63	& 0.712	& 0.73  \\
\multicolumn{2}{c|}{} & VAST & 11.11 & 0.56  & 0.58 & 14.6 & 0.566  & 0.61  \\ \midrule
\multicolumn{2}{c|}{Neural PLDA} & CMN2 & 11.33  & 0.609  & 0.62   & 10.06	& 0.699	& 0.711  \\
\multicolumn{2}{c|}{(BCE Loss, Random Init)} & VAST & 11.52 & 0.449 & 0.45 & 15.39 & 0.636  & 0.64 \\ \midrule
\multicolumn{2}{c|}{Neural PLDA} & CMN2 &11.04	& 0.564	& 0.58  & 08.97	&0.603	& 0.726  \\
\multicolumn{2}{c|}{(BCE Loss, Kaldi Init)} & VAST  & 07.41  &  0.416   & 0.527 &  14.60    & 0.578  &  0.627  \\ \midrule
\multicolumn{2}{c|}{Neural PLDA} & CMN2 &10.50	& \textbf{0.524}	& 0.532  & 09.78	& \textbf{0.598}	& 0.654 \\
\multicolumn{2}{c|}{(Soft detection cost)} & VAST & 07.41  &  \textbf{0.370} &   0.38  & 13.65   & \textbf{0.525}  &  0.585 \\ \midrule
\multicolumn{2}{c|}{Neural PLDA} & CMN2 & 11.20	& 0.540	& 0.562  & 10.23 & 0.646 & 0.678  \\
\multicolumn{2}{c|}{(Soft detection cost+0.1*BCE)} & VAST &11.11	& 0.374	& 0.389  &15.12	& 0.550	& 0.573  \\ \bottomrule
 \end{tabularx}

\label{sre:results}
\end{table*}

\section{Experiments}\label{sec:results}
We perform several experiments with the proposed neural net architecture and compare them with  various discriminative back-ends previously proposed in the literature such as the discriminative PLDA \cite{burget2011discriminatively} and pairwise Gaussian back-end~\cite{cumani2013pairwise}. We also compare the performance with the baseline system using Kaldi recipe that implements the generative PLDA model based scoring.

\pagebreak
For all the pairwise generative/discriminative models, we train the back-end using the trials sampled from previous NIST SRE evaluation sets along with randomly sampled target and non-target pairs which are matched by source and gender. We use about $5.3$ million trials for this training sampled from NIST SRE 04-10 as well as the NIST SRE16 trials. We also sample training data from Mixer-6 and Switchboard 1\&2 corpora. The evaluation of the models are performed on the telephone conditions (CMN2) and the video conditions (VAST) of the NIST SRE 2018 challenge. 

\subsection{Kaldi PLDA Baseline}
The primary baseline to benchmark our systems is the PLDA back-end implementation in the Kaldi toolkit. The Kaldi implementation models the average embedding x-vector of each training speaker. The x-vectors are centered, dimensionality reduced using LDA, followed by unit length normalization. By setting various dimensions, the best performance  on SRE 2018 development set was achieved with LDA dimension of $170$. The linear transformations and the Kaldi PLDA matrices are used to initialize the proposed pairwise PLDA network. 

\subsection{Discriminative PLDA (DPLDA)}
In \cite{burget2011discriminatively}, an expanded vector $\boldsymbol{\varphi}(\Eta_e,\Eta_t)$ representing a trial $(\Eta_e,\Eta_t)$ was computed using a quadratic kernel as follows: 
\begin{align}
\boldsymbol{\varphi}(\Eta_e,\Eta_t) = \begin{bmatrix}
    \text{vec}(\Eta_e\Eta_t\transpose + \Eta_t\Eta_e\transpose)\\
    \text{vec}(\Eta_e\Eta_e\transpose + \Eta_t\Eta_t\transpose)\\
    \text{vec}(\Eta_e + \Eta_t)\\
    1
    \end{bmatrix}
\end{align}
The PLDA log likelihood ratio score can be written as the dot product of a weight vector $\boldsymbol{w}$ and the expanded vector $\boldsymbol{\varphi}(\Eta_e,\Eta_t)$.
\begin{align}
    s = \boldsymbol{w}\transpose \boldsymbol{\varphi}(\Eta_e,\Eta_t)
\end{align}
We implemented the DPLDA in PyTorch by expanding the centered, LDA transformed and length normalized x-vectors from Kaldi baseline. Once the weight vector ${\boldsymbol w}$ is trained, the score on the test trials was performed using the inner product of the weight vector with the quadratic kernel.

\subsection{Pairwise Gaussian Back-end (GB)}
The Pairwise Gaussian Back-end \cite{ramoji2019LEAP,cumani2014generative}  models the pairs of enrollment and test x-vectors, $\Eta = [ \Eta_{1}\transpose \,\, \Eta_{2}\transpose ]\transpose$. The x-vector pairs are modeled using a Gaussian distribution  with parameters $(\Mu_t, \Sigma_t)$ for target trials while the non-target pairs are modeled by a Gaussian distribution with parameters $(\Mu_{nt}, \Sigma_{nt})$. These parameters are estimated by computing the sample mean and covariance matrices of the target and non-target trials in the training data. The log-likelihood ratio ($LLR$) for a new trial is then obtained as:
$$ LLR = -(\Eta-\Mu_t)\transpose\Sigma_t\inv(\Eta-\Mu_t) +(\Eta-\Mu_{nt})\transpose\Sigma_{nt}\inv(\Eta-\Mu_{nt})$$
The Gaussian Back-end is also trained on the same pairs of target and non-target x-vector trials, after centering, LDA and length normalization.

\pagebreak
\subsection{Neural PLDA}
We perform various experiments using the neural PLDA architecture with different initialization methods and loss functions. We also experiment with the role of batch size parameter, the learning rate as well as the choice of loss function in the optimization. The optimal parameter choices were based on the SRE 2018 development set. 

For the binary cross entropy (BCE) loss function and the soft detection cost functions, we need to apply the sigmoid function on the scores at different thresholds. In this work, we also parameterize the threshold value and let the network learn the threshold value to minimize the loss.

The soft detection cost function is highly sensitive to small changes in false alarm probability. Hence, all experiments were conducted with a large batch sizes of 4096/8192. The learning rate was initialized to $10^{-3}$ and halved each time the validation loss increased twice in a row.

\subsection {Discussion of Results}
The performance of the various back-end systems are reported in Table~\ref{sre:results}. The PLDA baseline generalized considerably well for both development and evaluation sets for both CMN2 and VAST sources. The Discriminative PLDA (DPLDA) is found to perform well on the VAST set, but it fails to generalize on CMN2 conditions. The Pairwise GB model also performs better than Kaldi's PLDA baseline on VAST dataset, which is in line with what was observed previously in \cite{ramoji2019LEAP}. 

The Neural PLDA model with random initialization of all parameters performed significantly better than the DPLDA model on the development set, and marginally better on the evaluation set. We hypothesize this to be a result of the network architecture which has fewer parameters, and hence fewer degrees of freedom than DPLDA model which results in better generalization. When the parameters are initialized with the Kaldi PLDA back-end parameters, the discriminative training further improves the performance on the dev set.

The soft detection cost function helps further reduce the $C_{min}$ and generalizes much better than using only the cross-entropy loss alone. We observe significant relative improvements over the PLDA Baseline of 10\% and 38\% in terms of $C_{min}$ on the SRE 2018 Development sets, respectively on CMN2 and VAST conditions. On the SRE 2018 Evaluation set, the proposed apporach yields relative improvements of 7\% and 23\% for CMN2 and VAST conditions.

\section{Summary and Conclusions}\label{sec:summary}
This paper presents a step in the direction of exploring discriminative models for the task of speaker verification. Discriminative models allow the construction of end-to-end systems. However, discriminative models tend to overfit to the training data. In our proposed model, we constrain the parameter set to have lesser degrees of freedom, in order to achieve better generalization. We also propose a task specific differentiable loss function which approximates the NIST SRE 2018 detection cost. 

It is important to note that unlike cross entropy loss, the NIST SRE detection cost gives significantly more importance to the false alarms. We also find that initializing the proposed neural PLDA model using generative model parameters allows the model to improve over the baseline system performance.

We observe considerable improvements and better generalization with our proposed approach. We could attribute this to the choice of architecture as well as the choice of loss functions.  

\section{Acknowledgements}
The authors would like to thank the Ministry of Human Resources Development (MHRD) of India and the Department of Science and Technology (DST) for their support. We would also like to thank Bhargavram Mysore and Anand Mohan for the valuable discussions and their help during the SRE 2018 and 2019 Evaluation.

\pagebreak
\bibliographystyle{IEEEbib}
\bibliography{strings,Template}

\end{document}